\documentclass[11pt,a4paper]{article}
\usepackage{url}
\usepackage{rotating}
\usepackage{secdot}
\usepackage{color}
\usepackage[dvipsnames]{xcolor}
\usepackage{booktabs}
\usepackage{amsmath,amsfonts,amssymb}
\usepackage{wrapfig}
\usepackage{subcaption}
\usepackage{adjustbox}
\usepackage{siunitx}
\usepackage[normalem]{ulem}
\usepackage{colortbl}

\usepackage{mathrsfs}

\usepackage{graphicx}

\newlength{\mylengthleft}
\newlength{\mylength}
\usepackage{tikz}

\usetikzlibrary{arrows,shapes,positioning}
\usetikzlibrary{decorations.markings}
\tikzstyle arrowstyle=[scale=1.8]
\tikzstyle directed=[postaction={decorate,decoration={markings,
    mark=at position .65 with {\arrow[arrowstyle]{stealth}}}}]
\tikzstyle reverse directed=[postaction={decorate,decoration={markings,
    mark=at position .65 with {\arrowreversed[arrowstyle]{stealth};}}}]

\begin{document}
\pagestyle{plain}
\setcounter{page}{1}

\begin{center}
{\Large \bf In Search of Dense Subgraphs:\\*[2mm]
	How Good is Greedy Peeling?}\\*[8mm]

{\large Naga V. C. Gudapati, Enrico Malaguti, Michele Monaci}\\*[8mm]

{\small \it
DEI, University of Bologna,
Viale Risorgimento 2,
I-40136 Bologna, Italy.\\
\rm\texttt{\{chaitanya.gudapati, enrico.malaguti, michele.monaci\}@unibo.it}}\\*[2mm]

\end{center}


\begin{abstract}
The problem of finding the densest subgraph in a given graph has several applications in graph mining, particularly in areas like social network analysis, 
protein and gene analyses etc. 
Depending on the application, finding dense subgraphs can be used to determine regions of high importance, similar characteristics or enhanced interaction. 
The densest subgraph extraction problem is a fundamentally a non-linear optimization problem. Nevertheless, it can be solved in polynomial time by an exact algorithm based on 
the iterative solution of a series of maximum flow sub-problems. Despite its polynomial time complexity, 
the computing time required by the exact algorithms on very large graphs could be prohibitive. 
Thus, to approach graphs with millions of vertices and edges, one has to resort to heuristic algorithms. We provide an efficient implementation of a greedy heuristic from the 
literature that is extremely fast and has some nice theoretical properties. We also introduce a new heurisitic algorithm that is built on top of the greedy and 
the exact methods. 
An extensive computational study is presented to evaluate the performance of various solution methods
on a benchmark composed of 86 instances taken from the literature.
This analysis shows that the proposed heuristic algorithm proved very effective on a large number of test instances, often providing either the optimal solution or near-optimal solution within short computing times.

\end{abstract}

\section{Introduction}\label{sec:intro}

A graph is a mathematical structure containing vertices and edges that is often used to 
represent different real-life scenarios. 
Besides very traditional applications in transportation, mapping, logistics, etc., graphs may also be used to describe many social, biological, financial, and 
technological systems. In these cases, vertices represent individuals, cells, proteins, components, etc., and edges represent some kind of interaction between the vertices.
As a  result, Graph Theory is one of the most extensively researched areas in computer science.

Graph networks that arise in real-life applications have edges which are either weighted or unweighted. While unweighted edges simply represent some connection 
between two vertices, weighted edges are used to indicate the importance of a connection in the graph, or the 
time required for travelling on a given edge, or probability of an edge to occur in the network. 
The edges could be further directed or undirected: the former models one-way relationships, like  the ``follow" network in Twitter,
while the latter are used for two-ways connections, for instance, Facebook friendships. 

One of the most interesting problems in social networks is the identification of dense areas.
Intuitively, dense areas in a graph can be considered to be a subset of highly-connected vertices that correspond to regions where there is more interaction among the vertices.
For instance, consider a network describing the interactions between various Internet service providers, exchange points, customers, and other 
related parties: identifying dense subgraphs in this network allows us to detect critical points of failure, which could further help in planning for contingencies 
to mitigate unplanned service outages. 
Similarly, for social networks, dense subgraphs identify areas of common interests and communities. Many other examples, where finding dense subgraphs is a key problem 
are detailed in \cite{LRJA10} and in \cite{F10}.

\section{Definition of the Problem}\label{sec:problem}

In this section we give a formal definition of the problem.
Let $G = (V, E)$ be an unweighted, undirected graph with vertex set $V$ and edge set $E$. Through out the text, we will assume that $G$ is a simple graph, i.e., there are
no multiple edges connecting the same pair of vertices.
The {\em density} of $G$, sometimes referred to as {\em average degree}, is defined as
\begin{equation}\label{eq:density}
f(G) = \frac{|E|}{|V|},
\end{equation}
and corresponds to the ratio between the number of edges and the number of vertices in the graph.

For a given subset of vertices $S \subseteq V$, we define $E(S)$ as the induced set of edges, i.e., $E(S) = \{e=(u,v) \in E: u \in S, v \in S\}$,
and $G(S) = (S, E(S))$ as the subgraph {\em induced} by $S$. When no confusion arises, we will write that set $S$ has a density
\begin{equation}\label{eq:notation}
	f(S) = f(G(S)) = \frac{|E(S)|}{|S|}
\end{equation}

Given an unweighted graph $G = (V, E)$, the {\em Densest Subgraph Extraction} (DSE) problem requires to determine a subset $S \subseteq V$ of vertices that induces a subgraph of maximum density.

As already mentioned, in many applications each edge $e \in E$ has a positive {\em weight} $w_e$, which could, for instance, be used to represent the importance of a relationship between
two vertices in the network. Weighted graphs can also be used to model a unique scenario where the actual edge set is unknown and each potential edge has associated a non-negative probability. In this probabilistic setting, one is interested in finding a subgraph that has a large
probability to be the one with maximum density. This leads to a natural extension of the density definition \eqref{eq:density} to the weighted case
density of a weighted graph as 
\begin{equation}\label{eq:wei_density}
f^{w}(G) = \frac{\displaystyle{\sum_{e \in E}} w_e}{|V|}.
\end{equation}
Similarly, we can define the weighted density for a given set $S \subseteq V$ of vertices.

The aforementioned density definitions are valid for undirected graphs only. For directed graphs, different definitions are typically used and
we refer the interested reader to \cite{C00, KS09}.

The DSE problem has been studied since the early 1980s. Though this problem is fundamentally an unconstrained non-linear optimization problem, it can still be solved
efficiently. Indeed, a flow-based algorithm to get an optimal solution of the problem for unweighted graphs was introduced in 
\cite{PQ82} and it requires utmost $|V|$ max-flow (min-cut) operations on a network of $|V|+2$ vertices i.e., it runs in polynomial time. 
Later, an alternative flow-based algorithm with better computational complexity was introduced in \cite{G84}. This algorithm
determines the densest subgraph in only  $\mathcal{O}(\log(|V|))$ max-flow operations and can be easily extended to weighted graphs.
Finally, a parametric max-flow algorithm which can solve the DSE with a single max-flow computation was given in 
\cite{GGT89}. This parametric max-flow algorithm improves upon the complexity of the previous method described in \cite{G84} by a factor of $\log(|V|)$ but that improvement in computational complexity can not be extended to weighted graphs.

Though solvable in polynomial time, computing densest subgraphs using flow-based algorithms could be very time consuming
for very large graphs. 
Thus, when real-world applications that involve millions of vertices and edges are considered, one has to resort to heuristics. 
One of the most important heuristic algorithms for the DSE problem is the {\em Greedy Peeling} introduced in \cite{AITT00}. 
Besides being very fast in practice, this algorithm has nice theoretical properties. It has been proved in \cite{C00}
that this algorithm has a worst-case 2-approximation, i.e., the density of the subgraph found by {\em Greedy Peeling} will be at least half of the density of the optimal subgraph.
Finally, we mention a variant of the {\em Greedy Peeling} algorithm, introduced in \cite{Bahmani:2012:DSS:2140436.2140442}, that can be implemented in a distributed way to reduce
the memory requirement for storing all the data. The resulting algorithm produces a solution with a worst-case approximation of $(2 + 2\epsilon)$ for any $\epsilon > 0$.

In some applications, additional constraints are imposed to limit (either from below or from above) the size of set $S$; in this case,
the resulting problem becomes an $\cal{NP}$-hard problem. 
An extensive discussion on finding dense subgraphs with size bounds can be found in \cite{AC09}.

Many alternative definitions of density have been proposed in the literature.
Indeed, the average degree may produce subgraphs that have a large number of vertices, and not extensively connected.  
For instance, a clique, which is intuitively a dense area in a graph, might not be the densest subgraph according to the average degree 
definition, as another larger and loosely connected subgraph could produce a bigger ratio according to \eqref{eq:density}.
Additional considerations about the downsides of using definition \eqref{eq:density} as a metric to find the dense subgraphs
are given in \cite{HS18}.
A different density metric, called {\em quasi-clique}, was introduced in \cite{TBGGT13}; according to
this definition, the density of graph $G = (V, E)$ is given by $f(G) = |E(S)| - \alpha\binom{|S|}{2}$, where $\alpha$ is a tuning 
parameter. This function tends to produce subgraphs that are relatively compact and well connected. The {\em discounted average degree}, proposed in \cite{HS18}, is defined as
$f(S) = \frac{|E(S)|}{|S|^{\beta}}$, where $\beta$ is a parameter that can be chosen to affect the size of the desired subgraph. 
The authors showed that this metric favours subgraphs that have certain desirable properties, like increased connectivity between vertices, and more compact subgraph.
Other than these two definitions, depending on the type of graph, there have been many other definitions of density, see, e.g., 
\cite{ARS02, BBH11, CS12}. 
While each of these metrics might give a more compact subgraph with respect to the average degree, there is no general consensus 
on the use of alternative metrics for finding dense subgraphs, and average degree remains the most common and accepted.
Besides the average degree metric, other metrics based on density have been considered, like edge ratio, triangle density, and triangle ratio
(see \cite{T14} for details).\\*[1ex]
The rest of the paper is organized as follows. 
In Section \ref{sec:algo} we will review the main exact and heuristic approaches from the literature for solving the DSE problem.
Section \ref{sec:hybrid} introduces a new {\em Hybrid Algorithm} that is built on top of {\em Greedy Peeling} and Exact Algorithms.
All algorithms are computationally  tested in Section \ref{sec:computational} on a large set of graph instances taken from the literature including both unweighted and weighted graphs. Finally, in Section \ref{sec:conclusions} we draw some conclusions.

This paper has three main contributions. From a practical viewpoint, we introduce a simple heuristic algorithm that is built on top of the greedy heuristic and any exact method.
Our proposed algorithm is typically very fast, produces solutions that improve over the greedy solution, and gives us near optimal solutions.

From a theoretical point of view, we present a simple graph instance where the {\em Greedy Peeling} algorithm approaches its worst-case performance.
To the best of our knowledge, there is only one contribution in the literature showing a similar behaviour of the {\em Greedy Peeling} algorithm (see \cite{TsourakakisKDD15}) but the example we present is simpler than the previous one.

Finally, from a computational study perspective, we present a through computational study that is by far the most extensive reported in the 
literature for this class of problems. While most of the previous works in the literature have dealt with small or medium-sized instances, 
in this paper we make a considerable step forward concerning the problem size by considering graphs with tens of millions of vertices and hundreds of millions of edges. 
Our computational study shows that the practical performance of the Greedy Peeling algorithm is much better than its theoretical guarantee, and that a further improvement 
can be achieved with limited computational effort.

\section{Algorithms}\label{sec:algo}
In this section we discuss solution approaches for the DSE problem proposed in the literature.
The next section describes two exact algorithmic approaches, while Section \ref{sec:peeling} presents a greedy heuristic and analyzes its theoretical performance.

\subsection{Exact algorithms}\label{sec:exact}

The first exact algorithm we consider is the {\em Goldberg's Algorithm} which has been introduced in \cite{G84} and is a relatively fast exact algorithm to compute the 
densest subgraph in a given graph $G$. The algorithm primarily works by iteratively solving a series of max-flow problems on an augmented graph $G'$ which is 
constructed from the original $G$. 
At each iteration, the algorithm ``guesses" the density value say $g$, defines an augmented graph $G'$ according to the current $g$ value,
and computes the maximum flow on $G'$. 
As the maximum density value will lie in the interval $[0, |E(S)|]$, a binary search approach is used to determine the optimal $g$ value. 
It is proved in \cite{G84} that, as the optimal $g$ value can only take a finite set of values, the number of iterations of the algorithm is bounded 
by $\mathcal{O}(\log(|V|))$.
There are many efficient algorithms for solving max-flow problem (see, e.g., \cite{HNSS18}). 
Using the Push-Relabel algorithm (see \cite{GT88}), the maximum flow problem can be solved in $\mathcal{O}(|V|^3)$ time, producing an overall 
$\mathcal{O}(\log(|V|) \, |V|^3)$ time complexity.

A completely different exact solution method has been proposed in \cite{C00}. This approach describes the DSE problem by means of a {\em Linear Programming} (LP) model,
that can be solved using any general-purpose LP solver.  The LP model can also be easily extended to the weighted case with minor modifications. 
The model has $|V| + |E|$ variables and two constraints per edge, i.e., its size is polynomial in the size of the input graph. 
Despite this, the constraint matrix of the formulation can be massive and the memory requirements to solve the model can be prohibitive
for large graph instances. Typically this
produces computational performances  that are worse than those of the flow-based algorithm {\em Goldberg's Algorithm} discussed above. 
However, the LP model provides a good foundation for  finding the densest subgraphs in directed graphs and its related proofs as discussed in \cite{C00}. 

\subsection{Greedy Peeling Algorithm}\label{sec:peeling}

For very large graph instances, the application of the exact algorithms described in the previous section may require large memory and long  computational times.  This is where 
heuristic approaches can be used for getting reasonably good solutions quickly. 
The heuristic algorithm described in this section produces subgraph whose density is usually close to the optimal one. 

As the objective of DSE is to find a subgraph with best average degree, the algorithm consists of starting with the initial graph and
removing, one at a time, the vertex with smallest degree in the current graph. 
The resulting algorithm, called {\em Greedy Peeling}, is described in Figure \ref{greedyPeeling} and can be naively implemented to run in $\mathcal{O}(n^2)$ on a graph $G = (V, E)$ 
with $n = |V|$ vertices and $m = |E|$ edges.
To prove the time complexity it is enough to observe that there are $n$ iterations; each iteration requires $\mathcal{O}(n)$ time to find the vertex $u$ with minimum degree 
with respect to the current subgraph (breaking ties arbitrarily), and another $\mathcal{O}(n)$ time to update the subgraph once $u$ has been removed.
A more efficient implementation can be obtained using a ``degree-lists'' data structure, in which a list is defined for each possible value of the degree of a vertex. All vertices with same degree are placed in the same list and lists are ordered by increasing degree.
Using this data structure, the determination of the next vertex $u$ to be removed can be done in constant time, taking an arbitrary vertex in the first non-empty list.
Since removing vertex $u$ decreases the degree of its neighbours by one unit, updating the graph (essentially data-lists) can be done  by moving each neighbour of $u$ from its current list
to the previous one (i.e., to the list with degree one less than current degree). Since the number of vertex movements among the lists is equal to the number of edges of $G$, the time complexity of the algorithm
is $\mathcal{O}(n + m)$. The results in this paper (see Section \ref{sec:computational}) correspond to this implementation of the algorithm.

\begin{figure}[htb!]
\centering
\fbox{\hspace{\mylengthleft}\parbox[t]{\mylength}{\vspace*{0.5ex}
{\bf Algorithm Greedy Peeling}($V, E$)\\*[1ex]
initialize: $n := |V|$, $S_n := V$;\\
{\bf for} $i = n$ {\bf to} $1$ {\bf do}\\
\hspace*{3ex} let $u$ be the smallest degree vertex in $G(S_{i})$;\\
\hspace*{3ex} $S_{i-1} := S_i \setminus \{u\}$;\\
{\bf endfor}\\
$S^{H} := \arg \max_{i = 1, \dots, n} f(S_{i})$;\\
{\bf return} $S^{H}$
}
}
\caption{Algorithm Greedy Peeling.}\label{greedyPeeling}
\end{figure}

The {\em Greedy Peeling} algorithm can be easily extended to the weighted case by selecting, at each iteration, vertex $u$ as the one having the minimum 
weighted sum of all the incident edges with respect to the current subgraph.
However, the linear time complexity of the algorithm is not preserved because the degree-lists data structure cannot be used for graphs with general weights. 
Using Fibonacci heaps to determine, at each iteration, the minimum weighted degree vertex, the algorithm runs in $\mathcal{O}(m + n\log(n))$, see \cite{C00}.
The degree-lists implementation could however be used to determine the weighted dense subgraphs, similar to the unweighted case, if weights are 
either integer numbers or are all scaled to integers. 
In both cases, there is a considerable worsening in the performances of the algorithm as the number of lists to be considered is bounded
by the maximum weight degree of all vertices, i.e., it is pseudo-polynomial in the size of the input (and is strongly dependent on the 
number of significative digits in the weight values, if these are not integer).
In  this paper, we use binary heaps to implement the {\em Greedy Peeling} algorithm for the weighted cases and report 
results for this implementation, which works for both rational and integer weights, and is very fast in practice even for large graphs.

\subsubsection{Worst-case Analysis}

The theoretical performance of the {\em Greedy Peeling} algorithm was analyzed in \cite{C00} (and in \cite{AITT00} for a constrained version of the DSE problem), where
the worst-case performance ratio of the algorithm was proved to be equal to $2$.
To the best of our knowledge, the only example for which the approximation is asymptotically tight has been given in \cite{TBGGT13}. As the given instance is somehow complicated and
for the sake of completeness, we report a simpler instance where the worst-case approximation is approached. 

\begin{figure}[ht]
\centering
\begin{tikzpicture}
  \draw (2,0) node{$\bullet$} node[right]{\tiny $t + 1$};
  \draw[line width=0.3mm] (2,0) arc (0:15:2) node{$\bullet$} node[right]{\tiny $t+2$} ;
  \draw[line width=0.3mm] (30:2) node{$\bullet$} node[right]{\tiny $t+3$} arc (30:45:2) node{$\bullet$} node[shift={(0.45,.10)}]{\tiny $t+4$} ;
  \draw[line width=0.3mm] (60:2) node{$\bullet$}  arc (60:75:2) node{$\bullet$} ;
  \draw[line width=0.3mm] (90:2) node{$\bullet$}  arc (90:105:2) node{$\bullet$};
 \draw[line width=0.3mm] (120:2) node{$\bullet$}  arc (120:135:2) node{$\bullet$};
  \draw[line width=0.3mm] (150:2) node{$\bullet$} arc (150:165:2) node{$\bullet$}  ;
  \draw[line width=0.3mm] (180:2) node{$\bullet$} node[shift={(-0.5,.05)}]{\tiny $t+i$}  arc (180:195:2) node{$\bullet$} node[shift={(-0.75,0)}]{\tiny $t+i+1$}  ;
  \draw[line width=0.3mm] (210:2) node{$\bullet$}  arc (210:225:2) node{$\bullet$}  ;
  \draw[line width=0.3mm] (240:2) node{$\bullet$}  arc (240:255:2) node{$\bullet$}  ;
  \draw[line width=0.3mm] (270:2) node{$\bullet$}  arc (270:285:2) node{$\bullet$}  ;
  \draw[dashed] (290:2) arc (290:325:2) ;
  \draw [line width=0.3mm] (330:2) node{$\bullet$} node[right]{\tiny $t + 2p-1$} arc (330:345:2) node{$\bullet$} node[right]{\tiny $t + 2p$} ;


\draw (0,0) node{$\bullet$} node[shift={(0.3,0)}]{\tiny $0$};
\draw [line width=0.3mm] (0,0) -- (0.86602540,	0.5) node{$\bullet$} node[shift={(0.2,0.1)}] {\tiny $1$};
\draw [line width=0.3mm](0,0) -- (0.50000000,	0.866025404) node{$\bullet$} node[shift={(0.15,0.225)}] {\tiny $2$};
\draw[line width=0.3mm](0,0) -- (0.00000000,	1) node{$\bullet$} node[shift={(0,0.25)}] {\tiny $3$};
\draw[line width=0.3mm](0,0) -- (-0.50000000,	0.866025404) node{$\bullet$} node[shift={(-0.20,0.225)}] {\tiny $4$};
\draw[line width=0.3mm](0,0) -- (-0.86602540,	0.5) node{$\bullet$} node[shift={(-0.25,0.15)}] {\tiny $5$};
\draw[line width=0.3mm](0,0) -- (-1.00000000,	0) node{$\bullet$} node[shift={(-0.25,0.0)}] {\tiny $6$};
\draw[line width=0.3mm](0,0) -- (-0.86602540,	-0.5) node{$\bullet$} node[shift={(-0.25,-.10)}] {\tiny $7$};
\draw[line width=0.3mm](0,0) -- (-0.50000000,	-0.866025404) node{$\bullet$} node[shift={(-0.20,-.20)}] {\tiny $8$};
\draw[line width=0.3mm](0,0) -- (0.00000000,	-1) node{$\bullet$} node[shift={(0,-0.25)}] {\tiny $9$};;
\draw[line width=0.3mm](0,0) -- (0.86602540,	-0.5) node{$\bullet$} node[shift={(0.2,-0.10)}] {\tiny $t$};
\draw[line width=0.3mm,dotted] (283:1.22) arc (283:317:1.22) ;
\end{tikzpicture}
\caption{Bad instance for the Greepy Peeling. The graph has $1 + t + 2p$ vertices and $t + p$ edges.} \label{fig:BadInstance}
\end{figure}
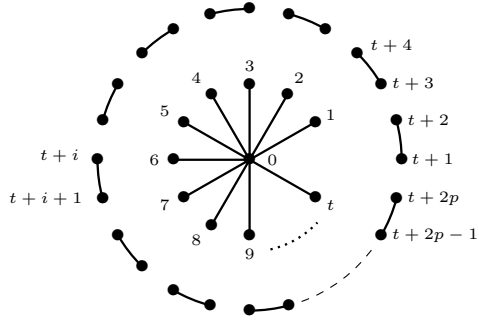

The bad instance, shown in Figure \ref{fig:BadInstance}, is a graph $G_{wc}$ which has an hub-and-spoke central structure with a single vertex acting as hub and a 
single vertex at the end of each of $t$ spokes. 
In addition to the hub-and-spoke arrangement of the central vertices, we have $2p$ vertices which are divided into $p$ pairs, each pair of vertices being connected by a single edge. 
It is trivial to see that the central hub-and-spoke structure, with $t$ edges and $t+1$ vertices is the densest subgraph in $G_{wc}$ and the optimal density is given by 
$f(S^*) = \frac{t}{t + 1}$, where $S^{*}$ denotes the set of vertices in the densest subgraph. 

Consider now the {\em Greedy Peeling} algorithm, and assume that ties are broken by selecting the vertex with minimum index among those with smallest degree.
Observe that each vertex in $G_{wc}$, except for the central one, has a degree of 1. Thus, the first selected vertex is $1$, then vertex $2$ follows, and so on.
At iteration $k$ the algorithm removes vertex $k$, and defines a subgraph with $1 + t + 2p - k$ vertices and $t + p - k$ edges, i.e., the density of the graph keeps decreasing.
Thus, the solution $S^G$ returned by the {\em Greedy Peeling} corresponds to the entire set of vertices and has a density $f(S^G) = \frac{t + p}{1 + t + 2p}$.

The ratio between the densities of the optimal solution and of the {\em Greedy Peeling} solution is is given by
$$
\frac{f(S^*)}{f(S^G)} = \frac{\frac{t}{t + 1}}{\frac{t + p}{1 + t + 2p}} = \frac{t (1 + t + 2p)}{(t + p) (t + 1)}
$$
For sufficiently large values of $p$ and $t$ with $p >> t$, the above ratio converges to $2$. 

\section{Hybrid algorithm}\label{sec:hybrid}

\begin{figure}[b!]
\centering
\fbox{\hspace{\mylengthleft}\parbox[t]{\mylength}{\vspace*{0.5ex}
{\bf Algorithm Hybrid}($V, E$)\\*[1ex]
{\scriptsize Peeling phase}\\*[1ex]
$S^1 := $ {\bf Greedy Peeling}($V, E$);\\*[1ex]
{\scriptsize Expansion phase}\\*[1ex]
$S^2 := \{v \in V: (u, v) \in E$ for some $v \in S^1\}$;\\*[1ex]
$E^2 := \{e=(u,v) \in E: u \in S^2, v \in S^2\}$;\\*[1ex]
{\scriptsize Exact phase}\\*[1ex]
$S^{H} := $ {\bf Exact}($S^2, E^2$);\\*[1ex]
{\bf return} $S^{H}$
}
}
\caption{Hybrid Algorithm.}\label{hybrid}
\end{figure}

In this section we present a {\em Hybrid Algorithm} that combines both {\em Greedy Peeling} and Exact Algorithm to improve the greedy solution value. 
The algorithm is given in Figure \ref{hybrid} and consists of three phases, namely Peeling phase, Expansion phase, and Exact phase. 
The first phase corresponds to the execution of the {\em Greedy Peeling} algorithm discussed in Section \ref{sec:peeling} and is intended to quickly produce an initial solution. 
Using the greedy solution, the Expansion phase obtains a ``core''  subgraph, which is likely to contain either all or most of the vertices in an optimal solution.
Finally, the Exact phase solves the DSE problem on the core  using an exact algorithm, for instance, the flow-based {\em Goldberg's Algorithm} or the LP approach described in Section \ref{sec:exact}.

The Expansion phase takes in input a subset of vertices $S^1$, possibly identified by the {\em Greedy Peeling}, expands the vertex set by adding all those vertices that 
are neighbors of one vertex in $S^1$, and defines the induced edge set $E^2$.
An implementation of this phase is described in Figure \ref{fig:expansion}.
Set $S^2$ is a list of all vertices that are currently included in the expanded graph. Before the expansion phase, $S^2 = \emptyset$. 
In the expansion phase we considers all vertices in $S^1$, one at a time.
For each $u \in S^1$, we consider all its neighbors; if the 
current neighbor $v$ is both is in $S^1 \cap S^2$, we add the edge $(u,v)$ to $E^2$. If $v \notin S^2$, we add vertex $v$ to $S^2$ 
and edge $(u, v)$ to $E^2$, and scan all neighbors of $v$; for each neighbor $k$ that is currently in set $S^2$ we also add an edge $(v, k)$ to $E^2$.

\begin{figure}[htb!]
\centering
\fbox{\hspace{\mylengthleft}\parbox[t]{\mylength}{\vspace*{0.5ex}
{\bf Procedure Expansion}($S^1, V, E$)\\
$S^2 := \emptyset$, $E^2 := \emptyset$;\\
{\scriptsize //consider each vertex $u$ in the input solution}\\*[1ex]
{\bf for each} $u \in S^1$ {\bf do}\\
\hspace*{3ex} $S^2 := S^2 \cup \{u\}$;\\
\hspace*{3ex} {\scriptsize add all neighbors of $u$}\\
\hspace*{3ex} {\bf for each} $v \in V: (u, v) \in E$ {\bf do}\\
\hspace*{6ex} {\bf if} $v \in S^1$ {\bf then}\\ 
\hspace*{9ex} {\bf if} $v \in S^2$ {\bf then} $E^2 := E^2 \cup \{(u, v)\}$;\\
\hspace*{6ex} {\bf else}\\
\hspace*{9ex} $S^2 := S^2 \cup \{v\}$, $E^2 := E^2 \cup \{(u, v)\}$;\\
\hspace*{9ex} {\scriptsize add edges between vertices that both are in $S^2 \setminus S^1$}\\
\hspace*{9ex} {\bf for each} $k \in S^2 \setminus S^1: (v, k) \in E$ {\bf do} $E^2 := E^2 \cup \{(v, k)\}$;\\
\hspace*{6ex} {\bf endif}\\
\hspace*{3ex} {\bf endfor}\\
{\bf endfor}\\
{\bf return} $(S^2, E^2)$
}
}
\caption{Expansion phase.}\label{fig:expansion}
\end{figure}

\begin{figure}[ht!]
\centering
\begin{tikzpicture}
    \coordinate (0) at (6.6, 3.5);
    \coordinate (1) at (5.3,4.00);
    \coordinate (2) at (6.5,0.9);
    \coordinate (3) at (6.2,-0.3);
    \coordinate (4) at (4,2.5);
    \coordinate (5) at (3,.8);
    \coordinate (6) at (5,0);
    \coordinate (7) at (5.3,1.2);
    \coordinate (8) at (3.7,-0.8);
    \coordinate (9) at (2,-0.1);
    \coordinate (10) at (4.8,-1.9);
    \coordinate (11) at (3.8,-2.0);

    \coordinate (14) at (15.6, 3.5);
    \coordinate (15) at (14.3,4.00);
    \coordinate (16) at (15.5,0.9);
    \coordinate (17) at (15.2,-0.3);
    \coordinate (18) at (13,2.5);
    \coordinate (19) at (12,.8);
    \coordinate (20) at (14,0);
    \coordinate (21) at (14.3,1.2);
    \coordinate (22) at (12.7,-0.8);
    \coordinate (23) at (11,-0.1);
    \coordinate (24) at (13.8,-1.9);
    \coordinate (25) at (12.8,-2.0);
    
    \coordinate (12) at ( 7.5, 1.2);
    \coordinate (13) at (10.5,1.2);

    \draw[thick] (0) -- (1);
    \draw[thick] (1) -- (4);
    \draw[thick] (1) -- (2) -- (7);
    \draw[thick] (2) -- (3) -- (6);
    \draw[line width=0.65mm, red, dashed] (4) -- (5) -- (6) -- (7) -- cycle;
    \draw[line width=0.65mm, red, dashed] (5) -- (7);
    \draw[line width=0.65mm, red, dashed] (4) -- (6);
    \draw[thick] (6) -- (8) -- (9) -- (5);
    \draw[thick] (8) -- (10);
    \draw[thick] (10) -- (11);
    
    \draw[thick] (14) -- (15);
    \draw[line width=0.65mm, red, dashed] (15) -- (18);
    \draw[line width=0.65mm, red, dashed] (15) -- (16) -- (21);
    \draw[line width=0.65mm, red, dashed] (16) -- (17) -- (20);
    \draw[line width=0.65mm, red, dashed] (18) -- (19) -- (20) -- (21) -- cycle;
    \draw[line width=0.65mm, red, dashed] (19) -- (21);
    \draw[line width=0.65mm, red, dashed] (18) -- (20);
    \draw[line width=0.65mm, red, dashed] (20) -- (22) -- (23) -- (19);
    \draw[thick] (22) -- (24);
    \draw[thick] (24) -- (25);

    \fill[black!20, draw=black, thick] (0) circle (3pt) node[black, above right] {\scriptsize{$1$}};
    \fill[black!20, draw=black, thick] (1) circle (3pt) node[black, above right] {\scriptsize{$2$}};
    \fill[black!20, draw=black, thick] (2) circle (3pt) node[black, above right] {\scriptsize{$3$}};
    \fill[black!20, draw=black, thick] (3) circle (3pt) node[black, above right] {\scriptsize{$4$}};
    \fill[red, draw=black, thick] (4) circle (3pt) node[black, above left] {\scriptsize{$5$}};
    \fill[red, draw=black, thick] (5) circle (3pt) node[black, above left] {\scriptsize{$8$}};
    \fill[red, draw=black, thick] (6) circle (3pt) node[black, below right] {\scriptsize{$7$}};
    \fill[red, draw=black, thick] (7) circle (3pt) node[black, above right] {\scriptsize{$6$}};
    \fill[black!20, draw=black, thick] (8) circle (3pt) node[black, above ] {\scriptsize{$9$}};
    \fill[black!20, draw=black, thick] (9) circle (3pt) node[black, above left] {\scriptsize{$10$}};
    \fill[black!20, draw=black, thick] (10) circle (3pt) node[black, above right] {\scriptsize{$11$}};
    \fill[black!20, draw=black, thick] (11) circle (3pt) node[black, above right] {\scriptsize{$12$}};
    
    \fill[black!20, draw=black, thick] (14) circle (3pt) node[black, above right] {\scriptsize{$1$}};
    \fill[red, draw=black, thick] (15) circle (3pt) node[black, above right] {\scriptsize{$2$}};
    \fill[red, draw=black, thick] (16) circle (3pt) node[black, above right] {\scriptsize{$3$}};
    \fill[red, draw=black, thick] (17) circle (3pt) node[black, above right] {\scriptsize{$4$}};
    \fill[red, draw=black, thick] (18) circle (3pt) node[black, above left] {\scriptsize{$5$}};
    \fill[red, draw=black, thick] (19) circle (3pt) node[black, above left] {\scriptsize{$8$}};
    \fill[red, draw=black, thick] (20) circle (3pt) node[black, below right] {\scriptsize{$7$}};
    \fill[red, draw=black, thick] (21) circle (3pt) node[black, above right] {\scriptsize{$6$}};
    \fill[red, draw=black, thick] (22) circle (3pt) node[black, above ] {\scriptsize{$9$}};
    \fill[red, draw=black, thick] (23) circle (3pt) node[black, above left] {\scriptsize{$10$}};
    \fill[black!20, draw=black, thick] (24) circle (3pt) node[black, above right] {\scriptsize{$11$}};
    \fill[black!20, draw=black, thick] (25) circle (3pt) node[black, above right] {\scriptsize{$12$}};
    
   \draw [thick, ->] (12) -- node [text width=2.5cm,midway,above,align=center ] {\bf expansion} (13);
    
   \node[text width=1cm] at (5.0,-2.8) {(a)};  
   \node[text width=1cm] at (14.0,-2.8) {(b)};
\end{tikzpicture}
\caption{an example of Expansion} \label{fig:ExpansionExample}
\end{figure}
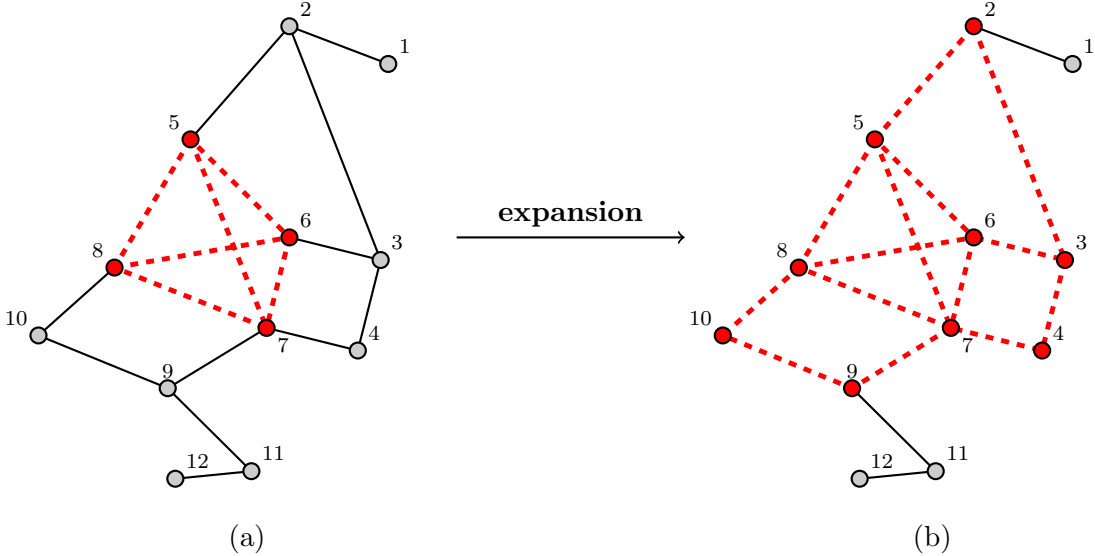

Figure \ref{fig:ExpansionExample} gives an example of the expansion phase. The original graph has 12 vertices and $S^1 = \{5, 6, 7, 8\}$. At first, $S^2 = \emptyset$.  
We can start at vertex $u = 5$ which makes $S^2 = \{$5$\}$ and consider the first of its neighbors, i.e., vertex $v = 2$. 
From the algorithm, we can add vertex $2$ to $S^2$ and the edge $(2,5)$ to $E^2$. Now, we scan the neighbors of $2$ and we can add an edge to $E^{2}$ if any of the 
neighboring vertices of $2$ are present in $S^{2}$. Since no new neighboring vertices of $2$ (essential vertex $1$) are present in $S^2$, we do not add any new edges to $E^{2}$. 
So we have $S^{2} = \{5,2\}$ and $E^{2} = \{(2,5)\}$. 
Then, we examine the other neighboring vertices of $5$ namely $6, 7$ and $8$. As all these vertices belong to $S^1$ and none of them are in $S^{2}$, no action is taken.
Then, we move on the next member in $S^{1}$, i.e. $u = 6$. We add $6$ to $S^{2}$  and examine the neighbors of $6$. We have vertex $v = 3$ that can be added to $S^{2}$ and the edge $(3,6)$ 
can be added to $E^{2}$. Now, $S^{2} = \{5,2,6,3\}$ and $S^{2} \setminus S^{1} = \{2 ,3\}$, implying that edge $(3,2)$ has to be added to $E^{2}$.
Since no other edge can be added, we then move on to the next neighbor of $6$, namely $5$. When considering this vertex, edge $(6,5)$ can be added to $E^{2}$ as $5$ is in 
both $S^{1}$ and $S^{2}$.
As all the neighbors of $6$ have been considered, we move onto the next vertex in $S^{1}$, i.e. $7$. 
We continue doing the above process for all the members in $S^{1}$ until we get the expanded subgraph, shown in Figure \ref{fig:ExpansionExample}(b)

In the third phase, an exact algorithm is applied to the graph obtained by expansion.
Typically this graph is much smaller than the original one, allowing a fast execution of the exact algorithm. In addition, 
if the flow-based algorithm is used, the greedy solution value, combined with the 2-approximation guarantee of the method, produces good initial lower and upper bounds for 
the value of the density, which can be used to speed up the binary search.
The biggest caveat is that there are instances for which the {\em Greedy Peeling} produces very large subgraphs. In this situation, the expansion procedure may require a very
long computing time, and often returns the complete graph, making the approach impractical.

\section{Computational experiments}\label{sec:computational}

\subsection{Setup and Programming}

All the algorithms described in this paper were implemented in C++ using standard containers, like, e.g., {\tt std::vector}, {\tt std::queue} etc. 
We used the GCC compiler with a high level of optimization enabled ({\tt -O3}). All our experiments were executed on a computer equipped with an Intel(R) Xeon(R) 
CPU E3-1220 V2 @ 3.10GHz CPU and 16 GB of RAM; all computing times given below are expressed in milliseconds.

In the following we report the results obtained using the three algorithms, namely:
\begin{itemize}
	\item The {\em Greedy Peeling} discussed in Section \ref{sec:peeling}.
	\item The {\em Hybrid Algorithm} of Section \ref{sec:hybrid}.
	\item The flow-based exact algorithm ({\em Goldberg's Algorithm}) of Section \ref{sec:exact}. This algorithm embeds a push-relabel algorithm to compute the max-flow 
	(min-cut) with $\mathcal{O}(|V|^3)$ time complexity.
	It should be noted that this algorithm requires to construct an augmented network which has twice the number of edges than the original graph network. 
	As a result, the augmented network could occupy very large space in memory, and hence the algorithm may fail for memory requirement on very large instances.
\end{itemize}	

We analyzed both weighted and unweighted instances (see below). For weighted problems, the 
{\em Greedy Peeling} was implemented using binary heaps as this solution turned out to be much more efficient than using the degree-list implementation.
As to the exact algorithm, it required very minor modifications for handling the weighted case as well.
The {\em Hybrid Algorithm} uses the {\em Greedy Peeling} and the {\em Goldberg's Algorithm} as modular components to create and solve the expanded subgraph respectively, 
while the expansion procedure clearly is not affected by the presence of weights on the edges.

\subsection{Testbed}
All the instances, both unweighted and weighted, were taken from Suite Sparse Collection \cite{sparse.tamu}.  
To select the instances, we considered all graphs that:
\begin{itemize}
	\item[(i)] are classified as {\em undirected graph} or {\em undirected weighted graph} or {\em undirected graph with communities} or {\em undirected random graph};
	\item[(ii)] have at least 20,000 vertices;
	\item[(iii)] have at most 65,000,000 vertices and 150,000,000 edges;
	\item[(iv)] have only positive weights (for weighted instances).
\end{itemize}

This produced a testbed with 170 instances. 
The benchmark includes 50 census-based weighted graphs (like xx2010 in Table \ref{tab:weighted}) that have very similar characteristics. 
To avoid presenting very similar results, we decided to consider only the ten largest among these instances. 
In addition, we have also considered three large directed graphs (called Wikipedia instances), that were present in the computational analysis in \cite{TsourakakisKDD15}; 
for these instances, minor modifications were required, e.g., converting directed arcs to undirected edges and removing duplicated edges.
Finally, we do not present the results on some graphs where the greedy algorithms fails. We explain the reason later in this section for this specific exclusion.  

Most of the available graphs are unweighted and hence we have partitioned them into different buckets, depending on their size. 
The {\bf Medium} bucket contains those instances which have less than 1,000,000  vertices. 
The {\bf Large} bucket contains instances having more than 1,000,000 vertices but less than 10 million vertices and less than 50,000,000 edges. 
Finally, the {\bf Xtra-Large} bucket includes all the remaining instances. 

\subsection{Analysis}
In this section we report the outcome of our computational experiments. The results are given using a table for each set of instances. 
The  bold numbers in the tables below indicate the best density found and which algorithm finds it for the first time.
The tables report, for each instance, the following information:
\begin{itemize}
	\item The name of the problem and the main characteristics of the graph.
	\item For the {\em Greedy Peeling} algorithm: the required computing time $T_G$ and the associated density value $f_G$.
	\item For the {\em Hybrid Algorithm}: the computing time for the expansion phase and for the exact phase ($T_2$ and $T_3$, respectively), 
	the overall computing time $T_H$ of the algorithm and the density value $f_H$ of the best solution found.
	\item For the Exact algorithm: the required computing time $T_E$ and the density value $f^*$.
\end{itemize}	
If an algorithm runs out of memory during its execution, we report the failure by `--'.

\begin{table}[h!]
	\footnotesize
\tabcolsep=2pt
\begin{adjustbox}{center}
\begin{tabular}{rrr|rr|rrrr|rr}
\hline
\multicolumn{3}{c|}{Graph Properties} & \multicolumn{2}{c|}{{\em Greedy Peeling}} & \multicolumn{4}{c|}{{\em Hybrid}} & \multicolumn{2}{c}{Exact}\\
\hline
          Problem &   $|V|$ &      $|E|$ &  $T_G$ &          $f_G$ &    $T_2$ &     $T_3$ &     $T_H$ &         $f_H$ &     $T_E$ &        $f^*$ \\
\hline
              144 & 144,649 &  1,074,393 &   53.7 &         7.4416 &  30114.8 &  259526.0 &  289694.5 &  {\bf 7.4559} &  280444.9 &       7.4559 \\
             598a & 110,971 &    741,934 &   37.6 &         6.8043 &   5066.7 &   35843.3 &   40947.6 &  {\bf 6.8792} &   73151.8 &       6.8792 \\
      as-22july06 &  22,963 &     48,436 &    3.2 &  {\bf 19.9423} &     11.4 &     737.0 &     751.5 &       19.9423 &    1317.4 &      19.9423 \\
             auto & 448,695 &  3,314,611 &  181.3 &         7.4495 &  89415.8 &  310410.2 &  400007.2 &        7.5211 &  622512.7 & {\bf 7.5213} \\
       ca-CondMat &  23,133 &     93,439 &    4.3 &        12.5000 &      0.5 &       8.6 &      13.3 & {\bf 13.3667} &    2336.1 &      13.3667 \\
 caidaRouterLevel & 192,244 &    609,066 &   50.2 &        25.5167 &      9.1 &     223.3 &     282.7 & {\bf 25.7750} &   23785.8 &      25.7750 \\
 citationCiteseer & 268,495 &  1,156,647 &   96.3 &        12.0019 &    205.8 &    6268.2 &    6570.3 & {\bf 12.1808} &   59115.4 &      12.1808 \\
coAuthorsCiteseer & 227,320 &    814,134 &   60.5 &  {\bf 43.0000} &      4.5 &      58.7 &     123.7 &       43.0000 &   25229.9 &      43.0000 \\
    coAuthorsDBLP & 299,067 &    977,676 &   78.1 &        57.0000 &      5.5 &      74.5 &     158.0 & {\bf 57.0690} &   35584.6 &      57.0690 \\
       com-Amazon & 334,863 &    925,872 &  104.4 &         3.8327 &   2163.0 &    8674.8 &   10942.2 &  {\bf 4.8041} &   53902.4 &       4.8041 \\
         com-DBLP & 317,080 &  1,049,866 &   94.5 &        56.5000 &      6.0 &      77.5 &     178.0 & {\bf 56.5652} &   38550.6 &      56.5652 \\
 coPapersCiteseer & 434,102 & 16,036,720 &  296.3 & {\bf 422.0000} &    229.9 &    3316.2 &    3842.4 &      422.0000 &  205527.8 &     422.0000 \\
     coPapersDBLP & 540,486 & 15,245,729 &  358.4 & {\bf 168.0000} &     67.8 &    2324.6 &    2752.8 &      168.0000 &  233183.4 &     168.0000 \\
              cs4 &  22,499 &     43,858 &    2.7 &         1.9493 &    247.4 &    8059.1 &    8309.2 &  {\bf 1.9526} &    9008.9 &       1.9526 \\
        dblp-2010 & 326,186 &    807,700 &   62.8 &  {\bf 37.0000} &      4.0 &      15.7 &      82.4 &       37.0000 &   26000.4 &      37.0000 \\
    delaunay\_n15 &  32,768 &     98,274 &    4.7 &   {\bf 2.9991} &    710.2 &   12997.1 &   13712.0 &        2.9991 &   14375.0 &       2.9991 \\
    delaunay\_n16 &  65,536 &    196,575 &   10.4 &   {\bf 2.9995} &   2835.5 &   50002.7 &   52848.5 &        2.9995 &   49406.5 &       2.9995 \\
    delaunay\_n17 & 131,072 &    393,176 &   23.0 &   {\bf 2.9997} &  11103.4 &  147668.5 &  158794.9 &        2.9997 &  145617.5 &       2.9997 \\
    delaunay\_n18 & 262,144 &    786,396 &   47.6 &   {\bf 2.9999} &  44140.1 &  394457.4 &  438645.1 &        2.9999 &  427411.9 &       2.9999 \\
    delaunay\_n19 & 524,288 &  1,572,823 &   96.6 &   {\bf 2.9999} & 176182.4 & 1740164.0 & 1916442.9 &        2.9999 & 1768799.0 &       2.9999 \\
     dictionary28 &  52,652 &     89,038 &    5.8 &  {\bf 12.5000} &      1.2 &       4.1 &      11.0 &       12.5000 &    2634.8 &      12.5000 \\
         fe\_body &  45,087 &    163,734 &    7.1 &         3.9043 &      2.4 &     159.0 &     168.5 &        3.9213 &    5421.1 & {\bf 4.0490} \\
        fe\_ocean & 143,437 &    409,593 &   22.7 &         2.8734 &   6533.3 &   49005.3 &   55561.3 &        2.8964 &   80359.5 & {\bf 2.8966} \\
        fe\_rotor &  99,617 &    662,431 &   27.5 &         6.6571 &  12459.8 &  146689.0 &  159176.4 &  {\bf 6.6920} &  159632.2 &       6.6920 \\
        fe\_tooth &  78,136 &    452,591 &   20.1 &         5.9171 &   2319.0 &   25546.7 &   27885.8 &        5.9778 &   58032.8 & {\bf 5.9801} \\
   loc-Brightkite &  58,228 &    214,078 &   11.0 &        40.5571 &     12.0 &     492.6 &     515.5 & {\bf 40.5591} &    6124.8 &      40.5591 \\
      loc-Gowalla & 196,591 &    950,327 &   62.1 &        43.8000 &    174.7 &   11902.8 &   12139.6 & {\bf 43.8018} &   32753.3 &      43.8018 \\
  luxembourg\_osm & 114,599 &    119,666 &   10.1 &         1.1548 &      2.5 &       2.6 &      15.2 &        1.2667 &    4338.7 & {\bf 1.5238} \\
             m14b & 214,765 &  1,679,018 &   79.9 &         7.8266 &  71078.1 &  185721.1 &  256879.2 &  {\bf 7.8694} &  238330.9 &       7.8694 \\
    mycielskian15 &  24,575 &  5,555,555 &  101.4 & {\bf 333.5567} &  30001.7 &   97961.5 &  128064.6 &      333.5567 &  107600.5 &     333.5567 \\
    mycielskian16 &  49,151 & 16,691,240 &  322.3 & {\bf 530.8705} & 175641.3 &  305244.9 &  481208.5 &      530.8705 &  344396.4 &     530.8705 \\
    mycielskian17 &  98,303 & 50,122,871 & 1092.1 & {\bf 845.8977} &       -- &        -- &        -- &            -- & 1165647.2 &     845.8977 \\
rgg\_n\_2\_15\_s0 &  32,768 &    160,240 &    6.9 &         7.5500 &      0.4 &       1.4 &       8.7 &        7.6522 &    3336.3 & {\bf 7.8947} \\
rgg\_n\_2\_16\_s0 &  65,536 &    342,127 &   16.8 &         7.6471 &      0.8 &       2.1 &      19.6 &   {\bf 9.000} &    7824.2 &       9.0000 \\
rgg\_n\_2\_17\_s0 & 131,072 &    728,753 &   39.9 &         8.0000 &      1.6 &       1.4 &      42.8 &        8.2083 &   20552.0 & {\bf 8.9200} \\
rgg\_n\_2\_18\_s0 & 262,144 &  1,547,283 &   87.0 &        10.0769 &      3.1 &       2.9 &      93.0 & {\bf 10.4242} &   45015.0 &      10.4242 \\
rgg\_n\_2\_19\_s0 & 524,288 &  3,269,766 &  190.4 &         8.9474 &      5.6 &       1.5 &     197.6 & {\bf 10.1667} &  125960.6 &      10.1667 \\
             t60k &  60,005 &     89,440 &    5.9 &         1.4905 &   1036.3 &   91590.4 &   92632.6 &  {\bf 1.4914} &   83854.0 &       1.4914 \\
          usroads & 129,164 &    165,435 &   19.1 &         1.5789 &      1.7 &       0.9 &      21.6 &        1.6250 &   11992.8 & {\bf 1.7528} \\
       usroads-48 & 126,146 &    161,950 &   18.6 &         1.5714 &      2.6 &       1.0 &      22.2 &        1.6250 &   14238.7 & {\bf 1.7528} \\
             wing &  62,032 &    121,544 &    9.5 &         1.9596 &   1897.6 &   46318.4 &   48225.5 &  {\bf 1.9627} &   53894.6 &       1.9627 \\
\hline
\end{tabular}
\end{adjustbox}{}
\caption{Results on {\bf Medium} size instances.}
\label{tab:medium}
\end{table}

The results in Table \ref{tab:medium} show that the exact algorithm can handle quite efficiently instances of medium size: the required computing time is equal to 
137 seconds on average and no failure was experienced due to memory reasons.
The {\em Greedy Peeling} algorithm, though having a worst-case performance ratio equal to 2,
gives a very tight approximation on the optimal density in practice. 
The average gap with respect to the optimal density is below 3\% and the average CPU time is around 0.07 seconds. 
Nevertheless, the {\em Hybrid Algorithm} improves over the greedy solution in 27 cases (out of 41) and produces an average gap to the optimum of about 1\%. 
However, there are a number of instances for which the {\em Hybrid Algorithm} performs poorly in terms of computing time. 
We can also see that for the instance mycielskian17, the {\em Hybrid Algorithm} runs out of memory.
In Table \ref{tab:bad} we report all the instances where the ratio of $\frac{T_{H}}{T_{E}}> 0.75$ and where the {\em Hybrid Algorithm} runs out of memory. 
For each such instance, the table gives:
\begin{itemize}
	\item The name of the problem.
	\item The number of vertices $|V_G|$ in the subgraph produced by {\em Greedy Peeling} and the ratio between $|V_G|$ and the total number of vertices $V$.
	\item The number of vertices and edges ($|V_2|$ and $|E_2|$, respectively) in the expanded subgraph and the ratio between $|V_2|$ and the total number of vertices $V$.
\end{itemize}

\begin{table}[htb!]
\footnotesize
\tabcolsep=4pt
\begin{adjustbox}{center}
\begin{tabular}{r|rr|rrr}
\hline
\multicolumn{1}{c|}{Graph Properties} & \multicolumn{2}{c|}{{\em Greedy Peeling}} & \multicolumn{3}{c}{{\em Hybrid}} \\
\hline
      Problem &   $|V_G|$ & $|V_G|/|V|$ &   $|V_2|$ &      $E_2$ & $|V_2|/|V|$ \\
\hline                                                            
          144 &   144,649 &      0.9509 &   138,830 &  1,032,694 &      0.9598 \\
          cs4 &    22,499 &      1.0000 &    22,499 &     43,858 &      1.0000 \\
delaunay\_n15 &    32,768 &      1.0000 &    32,768 &     98,274 &      1.0000 \\
delaunay\_n16 &    65,536 &      1.0000 &    65,536 &    196,575 &      1.0000 \\
delaunay\_n17 &   131,072 &      1.0000 &   131,072 &    393,176 &      1.0000 \\
delaunay\_n18 &   262,144 &      1.0000 &   262,144 &    786,396 &      1.0000 \\
delaunay\_n19 &   524,288 &      1.0000 &   524,288 &  1,572,823 &      1.0000 \\
    fe\_ocean &   143,437 &      0.7212 &   109,860 &    315,384 &      0.7659 \\
    fe\_rotor &    99,617 &      0.9859 &    98,971 &    658,472 &      0.9935 \\
         m14b &   214,765 &      0.9634 &   210,693 &  1,647,651 &      0.9810 \\
           M6 & 3,501,776 &      0.9739 & 3,412,415 & 10,233,983 &      0.9745 \\
mycielskian15 &    24,575 &      0.3694 &    24,575 &  5,555,555 &      1.0000 \\
mycielskian16 &    49,151 &      0.3344 &    49,151 & 16,691,240 &      1.0000 \\
mycielskian17 &    98,303 &      0.2899 &    98,303 & 50,122,871 &      1.0000 \\
         t60k &    60,005 &      0.9977 &    59,935 &     89,313 &      0.9988 \\
         wing &    62,032 &      0.9971 &    61,994 &    121,461 &      0.9994 \\
\hline
\end{tabular}
\end{adjustbox}
\caption{Instances for which the {\em Hybrid Algorithm} can take a very long time.}
\label{tab:bad}
\end{table}

Table \ref{tab:bad} shows that the pitfalls of {\em Hybrid Algorithm} occur in those instances where the solution produced by {\em Greedy Peeling} has almost the same number
of vertices as the whole graph. And sometimes, even when {\em Greedy Peeling} produces a smaller and more compact solution, the expansion phase produces either the original graph or 
almost the original graph.  
In these cases, the expansion phase may be time consuming, and the application of the exact algorithm requires the same computing time
as solving the initial problem to optimality. Thus, the {\em Hybrid Algorithm} may overall be even slower than the direct application of the exact algorithm on the initial graph. 
The average computing time taken by the {\em Hybrid Algorithm} for the {\bf Medium} instances is around 115 seconds; if we exclude the pathological cases listed in Table \ref{tab:bad}, 
the hybrid time falls to around 21 seconds.

\begin{table}[htb!]
\footnotesize
\tabcolsep=2pt
\begin{adjustbox}{center}
\begin{tabular}{rrr|rr|rrrr|rr}
\hline
\multicolumn{3}{c|}{Graph Properties} & \multicolumn{2}{c|}{{\em Greedy Peeling}} & \multicolumn{4}{c|}{{\em Hybrid}} & \multicolumn{2}{c}{Exact}\\
\hline
                  Problem &      $|V|$ &      $|E|$ &  $T_G$ &    $f_G$ &     $T_2$ &     $T_3$ &     $T_H$ &          $f_H$ &       $T_E$ &         $f^*$ \\
\hline
               as-Skitter &  1,696,415 & 11,095,298 &  822.6 &  89.1810 &    2303.8 &   37273.4 &   40399.8 &  {\bf 89.4009} &    388513.9 &       89.4009 \\
                asia\_osm & 11,950,757 & 12,711,603 & 1342.2 &   1.7778 &     135.4 &       0.4 &    1478.1 &         1.7778 &    703145.0 &  {\bf 1.8513} \\
             belgium\_osm &  1,441,295 &  1,549,970 &  185.2 &   1.6000 &      15.4 &       0.1 &     200.6 &         1.6000 &     77872.7 &  {\bf 1.6750} \\
          com-LiveJournal &  3,997,962 & 34,681,189 & 3129.7 & 190.9845 &      82.8 &     695.2 &    3907.7 & {\bf 193.5136} &   1226155.5 &      193.5136 \\
              com-Youtube &  1,134,890 &  2,987,624 &  341.3 &  45.5778 &    1608.5 &   60642.4 &   62592.3 &  {\bf 45.5988} &    157970.6 &       45.5988 \\
             germany\_osm & 11,548,845 & 12,369,181 & 1734.4 &   1.6250 &     133.9 &       0.6 &    1868.9 &         1.6667 &    784833.6 &  {\bf 1.7500} \\
       great-britain\_osm &  7,733,822 &  8,156,517 & 1039.0 &   1.8710 &      93.7 &       1.3 &    1134.0 &   {\bf 1.9583} &    465254.6 &        1.9583 \\
               italy\_osm &  6,686,493 &  7,013,978 &  743.4 &   1.6250 &      80.2 &       0.4 &     824.0 &         1.6667 &    365157.8 &  {\bf 1.7778} \\
         netherlands\_osm &  2,216,688 &  2,441,238 &  298.8 &   1.6667 &      29.4 &       0.3 &     328.5 &   {\bf 1.7143} &    190545.5 &        1.7143 \\
 packing-500x100x100-b050 &  2,145,852 & 17,488,243 &  640.9 &   8.5361 &  147977.6 &  612576.4 &  761195.0 &         8.7361 &   2931714.4 &  {\bf 8.8078} \\
        rgg\_n\_2\_20\_s0 &  1,048,576 &  6,891,620 &  415.2 &  11.1212 &      14.1 &       4.4 &     433.7 &        11.6250 &    276226.5 & {\bf 11.6346} \\
        rgg\_n\_2\_21\_s0 &  2,097,152 & 14,487,995 &  930.6 &   9.3934 &      22.3 &       7.9 &     960.8 &  {\bf 11.9048} &    667290.4 &       11.9048 \\
        rgg\_n\_2\_22\_s0 &  4,194,304 & 30,359,198 & 1937.8 &  10.5503 &      58.4 &      25.0 &    2021.1 &   {\bf 12.550} &   1806177.7 &       12.5500 \\
            road\_central & 14,081,816 & 16,933,413 & 3285.6 &   1.6002 &     179.3 &      20.9 &    3485.8 &         1.7750 &   6231763.4 &  {\bf 1.9029} \\
               roadNet-CA &  1,971,281 &  2,766,607 &  315.0 &   1.6743 &      48.7 &     233.5 &     597.3 &   {\bf 1.9677} &    313535.2 &        1.9677 \\
               roadNet-PA &  1,090,920 &  1,541,898 &  177.7 &   1.6441 &      14.0 &      14.2 &     205.8 &         1.8571 &    234657.2 &  {\bf 1.8783} \\
               roadNet-TX &  1,393,383 &  1,921,660 &  216.8 &   1.7656 &      17.1 &       7.3 &     241.3 &   {\bf 2.0769} &     82250.8 &        2.0769 \\
            venturiLevel3 &  4,026,819 &  8,054,237 &  672.9 &   2.0014 & 1001420.7 & 111528.39 & 1113531.6 &   {\bf 2.0613} & 351929.5074 &        2.0613 \\
       wikipedia-20051105 &  1,634,989 & 18,540,603 & 1561.8 & 126.5925 &   14248.1 &  418588.5 &  434379.8 & {\bf 127.0162} &    872899.2 &      127.0162 \\
       wikipedia-20060925 &  2,983,494 & 35,048,116 & 3643.1 & 138.7406 &   43194.7 &  967617.4 & 1014476.4 & {\bf 140.5966} &   1919124.9 &      140.5966 \\
       wikipedia-20061104 &  3,148,440 & 37,043,458 & 3862.4 & 140.5598 &   47102.8 & 1031432.3 & 1082416.4 & {\bf 141.6711} &   2063044.9 &      141.6711 \\
         \hline
\end{tabular}
\end{adjustbox}
\caption{Results on {\bf Large} instances.}
\label{tab:large}
\end{table}

In Table \ref{tab:large} we present the results of our experiments on {\bf Large} instances. 
Based on the outcome of the results in Table \ref{tab:bad}, we do not run the {\em Hybrid algorithm} for those instances where the greedy solution (or the expanded subgraph) is almost as large as 
the complete original graph. In particular, we removed the graphs for which $\frac{|V_2|}{|V|} > 0.85$, namely the instances in the series delaunay series, hugebubbles, hugetrace, 
and hugetric, 
as well as instances 333SP, adaptive, AS365, channel-500x100x100-b050, NACA0015, and NLR. 
Note that $|V_2|$ can be computed in negligible time before performing the expansion, simply scanning all the edges that are incident to vertices in the greedy solution.

Results in Table \ref{tab:large} show that, like medium instances, {\em Hybrid Algorithm} consistently improves upon the density value produced by {\em Greedy Peeling}, 
and often times finds the optimal solution.  The {\em Hybrid Algorithm} was able to find the optimal solution in 13 cases out of 21 instances, and in 12 cases 
it was faster than the {\em Goldberg's Algorithm}. As for the remaining 8 instances that are not solved to optimality, the associated average gap is around 4\%. 
The average gap over all the 21 instances is around 1\%, much smaller than that of the {\em Greedy Peeling}, which is around 7\%. 
As for the computing time, {\em Greedy Peeling} just takes around 1.4 seconds on average, while the {\em Hybrid Algorithm} takes 215 seconds on average. 
The {\em Goldberg's Algorithm} takes almost 1100 seconds on average for solving these problems to optimality. 

In Table \ref{tab:Xtra-Large}, we present the results of the three algorithms for the eight {\bf Xtra-Large} instances in our benchmark. 
These graph instances were derived from real-life applications like gene networks (kmer series), road networks, social networks etc.
It can be immediately seen that the exact algorithm fails for all the instances due to memory limitation. 
For these instances, {\em Greedy Peeling} finds a dense subgraph within 10 seconds on average, despite running on some graphs having 
tens of millions of vertices and hundreds of millions of edges. 
The {\em Hybrid Algorithm} consistently improves upon the greedy solution for almost all these istances as well, the only exception being problem soc-orkut, for which the algorithm runs out of memory. 
Ignoring this instance, the average computing time taken by the {\em Hybrid Algorithm} is around 40 seconds, and 
the average improvement produced by this algorithm over the {\em Greedy Peeling} is by 9\%. 

\begin{table}[h!]
	\footnotesize
\tabcolsep=2pt
\begin{adjustbox}{center}
\begin{tabular}{rrr|rr|rrrr|rr}
\hline
\multicolumn{3}{c|}{Graph Properties} & \multicolumn{2}{c|}{{\em Greedy Peeling}} & \multicolumn{4}{c|}{{\em Hybrid}} & \multicolumn{2}{c}{Exact} \\
\hline
          Problem &      $|V|$ &       $|E|$ &   $T_G$ &           $f_G$ &   $T_2$ &    $T_3$ &    $T_H$ &         $f_H$ & $T_E$ & $f^*$ \\
\hline                                                                                               
      europe\_osm & 50,912,018 &  54,054,660 &  6869.1 &          1.7047 &   640.2 &     26.1 &   7535.4 &  {\bf 2.0000} &    -- &    -- \\
   hollywood-2009 &  1,139,905 &  56,375,711 &  1895.3 & {\bf 1104.0000} & 14712.7 & 199468.9 & 216076.9 &     1104.0000 &    -- &    -- \\
        kmer\_U1a & 67,716,231 &  69,389,281 & 26907.0 &          4.0000 &   862.0 &      2.0 &  27771.0 &  {\bf 4.0455} &    -- &    -- \\
        kmer\_V2a & 55,042,369 &  58,608,800 & 20570.0 &          6.9000 &   691.0 &     10.8 &  21271.8 &  {\bf 7.0909} &    -- &    -- \\
rgg\_n\_2\_23\_s0 &  8,388,608 &  63,501,393 &  4072.6 &         11.0476 &   112.6 &     25.6 &   4210.8 & {\bf 13.4000} &    -- &    -- \\
rgg\_n\_2\_24\_s0 & 16,777,216 & 132,557,200 &  8571.6 &         12.1220 &   237.8 &     15.5 &   8824.9 & {\bf 13.7143} &    -- &    -- \\
        road\_usa & 23,947,347 &  28,854,312 &  4545.7 &          1.5974 &   301.5 &      2.7 &   4849.9 &  {\bf 1.8462} &    -- &    -- \\
        soc-orkut &  4,847,571 & 106,349,209 &  9111.0 &        206.9307 &   509.7 &       -- &       -- &            -- &    -- &    -- \\
\hline
\end{tabular}
\end{adjustbox}
\caption{Results on {\bf Xtra-Large} instances.}
\label{tab:Xtra-Large}
\end{table}

Finally, Table \ref{tab:weighted} presents the results of the three algorithms on the weighted instances. 

The {\em Greedy Peeling} performs very well, and determines an optimal solution in 9 out of 16 cases; for instance mawi\_201512020000 it produces the same density value as the 
{\em Hybrid}, but optimality of the solution cannot be confirmed as the exact method failed.
The {\em Hybrid} improves over the greedy solution in 5 of the 6 remaining instances, in all these cases finding an optimal solution.
On average, the {\em Greedy Peeling} takes 2 seconds, while the {\em Hybrid Algorithm} takes 3 seconds. On the other hand, the exact algorithm requires almost 98 seconds to find the optimal solution. 
By removing the two mawi instances, we see that the average error of  the {\em Greedy Peeling} is around $1.7\%$, reduced to less than $0.02\%$ by  the {\em Hybrid Algorithm}.

\begin{table}[h!]
  \footnotesize
\tabcolsep=2pt
\begin{adjustbox}{center}
\begin{tabular}{rrr|rr|rrrr|rr}
\hline
\multicolumn{3}{c|}{Graph Properties} & \multicolumn{2}{c|}{{\em Greedy Peeling}} & \multicolumn{4}{c|}{{\em Hybrid}} & \multicolumn{2}{c}{Exact}\\
\hline
           Problem &      $|V|$ &      $|E|$ &   $T_G$ &             $f_G$ &     $T_2$ &    $T_3$ &    $T_H$              $f_H$ &    $T_E$ &            $f^*$ \\
\hline                                                                                                       
            ca2010 &    710,145 &  1,744,683 &   836.1 &  {\bf 6234021.00} &      21.1 &      0.5 &    889.9 &       6234021.00 & 102852.4 &       6234021.00 \\
     cond-mat-2003 &     31,163 &    120,029 &    26.3 &       {\bf 17.60} &       0.8 &      0.2 &     17.3 &            17.60 &   2993.6 &            17.60 \\
     cond-mat-2005 &     40,421 &    175,693 &    21.6 &       {\bf 23.00} &       1.1 &      0.0 &     24.5 &            23.00 &   4823.1 &            23.00 \\
            fl2010 &    484,481 &  1,173,147 &   482.8 &        3753682.46 &      14.7 &     19.9 &    544.5 & {\bf 3992056.53} &  59491.5 &       3992056.53 \\
            ga2010 &    291,086 &    709,028 &   285.2 &  {\bf 3929610.00} &       8.6 &      3.9 &    278.9 &       3929610.00 &  32925.6 &       3929610.00 \\
      human\_gene1 &     22,283 & 12,323,680 &   277.7 &       {\bf 62.67} &   26151.3 & 142179.7 & 168647.0 &            62.67 & 275029.7 &            62.67 \\
            il2010 &    451,554 &  1,082,232 &   519.4 &  {\bf 5508363.60} &      13.2 &      0.6 &    488.6 &       5508363.60 &  56272.4 &       5508363.60 \\
            mi2010 &    329,885 &    789,045 &   275.4 &        6993878.84 &      10.0 &      2.8 &    316.4 &       7370921.58 &  39722.7 & {\bf 7390000.23} \\
            mo2010 &    343,565 &    828,284 &   342.9 &  {\bf 1666117.50} &      10.6 &      0.5 &    342.3 &       1666117.50 &  41242.4 &       1666117.50 \\
mawi\_201512012345 & 18,571,154 & 19,020,160 &  8500.7 &         798116.43 &     557.3 &    119.7 &   9817.0 &  {\bf 927951.00} &       -- &               -- \\
mawi\_201512020000 & 35,991,342 & 37,242,710 & 16670.7 &  {\bf 1770103.00} & 1073.8856 &    183.5 &  19113.4 &       1770103.00 &       -- &               -- \\
       mouse\_gene &     45,101 & 14,461,095 &   423.5 &             27.75 &   34124.6 & 217645.7 & 252196.7 &      {\bf 28.47} & 504904.1 &            28.47 \\
            ny2010 &    350,169 &    854,772 &   316.5 &        2986674.11 &      10.9 &      2.4 &    349.7 & {\bf 3289936.62} &  42967.6 &       3289936.62 \\
            oh2010 &    365,344 &    884,120 &   340.1 &  {\bf 3826971.80} &      10.9 &      3.6 &    370.5 &       3826971.80 &  43120.3 &       3826971.80 \\
            pa2010 &    421,545 &  1,029,231 &   429.7 &  {\bf 3202713.00} &      12.1 &      0.4 &    429.4 &       3202713.00 &  52429.2 &       3202713.00 \\
            tx2010 &    914,231 &  2,228,136 &  1181.9 &        6563105.33 &      26.7 &      2.2 &   1225.0 & {\bf 6630141.80} & 119992.3 &       6630141.80 \\
\hline
\end{tabular}
\end{adjustbox}
\caption{Results on Weighted instances.}
\label{tab:weighted}
\end{table}

\section{Conclusions} \label{sec:conclusions}
In this paper, we have studied a non-linear graph optimization problem that requires to determine the densest subgraph in a 
given graph. We introduced a new heuristic algorithm that combines a fast and effective greedy algorithm and an exact method from the literature. 
We have provided  a simple instance for which the greedy algorithm shows its worst case performance.  
We have presented an efficient implementation of the algorithms to solve both unweighted and weighted problems, with the aim of attacking instances of very large size, like those
arising, e.g., in social network applications. To the best of our knowledge, this is the most comprehensive computational study on Densest Subgraph Extraction problem involving instances with tens of millions of 
vertices and hundreds of millions of edges.

\section*{Acknowledgments}
This research was supported by ``Mixed-Integer Non Linear Optimisation: Algorithms and Application'' consortium,  which has received funding from the European Union’s EU Framework Programme for Research and Innovation Horizon 2020 under the Marie Sk{\l}odowska-Curie Actions Grant Agreement No 764759.
\bibliography{biblio}
\bibliographystyle{plain}

\end{document}